\documentclass[a4paper]{article}

\usepackage{INTERSPEECH2019}
\usepackage[implicit=false]{hyperref}
\usepackage{subfigure}

\title{Towards Fine-Grained Prosody Control for Non-parallel Voice Conversion}
\name{Zheng Lian$^{1,2}$, Zhengqi Wen$^{1}$}
\address{
$^{1}$National Laboratory of Pattern Recognition, CASIA, Beijing, China\\
$^{2}$School of Artificial Intelligence, University of Chinese Academy of Sciences, Beijing, China}
\email{\{zheng.lian, zqwen\}@nlpr.ia.ac.cn}

\begin{document}
	
\maketitle
\begin{abstract}
In a typical voice conversion system, previous works utilized various acoustic features (such as the pitch, voiced/unvoiced flag and aperiodicity) of the source speech to control the prosody of converted speech. However, prosody is related with many factors, such as the intonation, stress and rhythm. It is a challenging task to perfectly describe prosody through handcrafted acoustic features. To address these difficulties, we propose to use prosody embeddings to describe prosody. These embeddings are learned from the source speech in an unsupervised manner. To verify the effectiveness of our proposed method, we conduct experiments on the popular benchmark database CMU-ARCTIC, and our Mandarin corpus. Experimental results show that our proposed method can improve the speech quality and speaker similarity of the converted speech. What's more, we observe that our method can even achieve promising results in singing conditions.
\end{abstract}

\section{Introduction}
Voice conversion (VC) aims to modify the source speaker's voice to sound like that of the target speaker while keeping the linguistic content unchanged. VC is an important research topic due to its wide applications, such as development of personalized speaking aids for speech-impaired subjects \cite{kain2007improving, biadsy2019parrotron, wang2020end}, novel vocal effects of singing voices \cite{luo2020singing, deng2020pitchnet}, and a voice changer to generate various types of expressive speech \cite{turk2010evaluation, kim2020emotional}.

The conventional voice conversion approach usually needs parallel training data, which contains pairs of the same transcription utterances spoken by different speakers \cite{stylianou1998continuous, takashima2013exemplar, desai2010spectral}. However, building such parallel data corpus is a highly expensive task. To address this problem, researches have proposed some methods for a voice conversion system that does not require the parallel training data \cite{hsu2016voice, huang2018voice, kaneko2018cyclegan}. Recently, phonetic posteriorgrams (PPGs) have been successfully applied to non-parallel VC and achieved both high naturalness and high speaker similarity of the converted speech \cite{sun2016phonetic, zhou2019cross}. PPG is a sequence of frame-level linguistic information representation obtained from the speaker-independent automatic speech recognition (SI-ASR) system. The PPGs based VC frameworks mainly have two key components: the conversion model and the vocoder. The conversion model converts PPGs extracted from the source speech into acoustic features of the target speaker. Then the vocoder uses these converted features to synthesize the speech waveform of the target speaker. However, we observe two limitations with existing works \cite{lu2019compact, tian2019speaker}.

Firstly, earlier works \cite{lu2019compact,tian2019speaker} mainly utilized acoustic features (such as the pitch, voiced/unvoiced flag and aperiodicity) of the source speech to control the prosody of the converted speech. However, prosody is related with many factors, such as the intonation, stress and rhythm. It is a challenging task to perfectly describe the prosody through acoustic features. The same challenge also exists in Text-To-Speech (TTS). To address this challenge, recent works learn to describe prosody that do not require explicit annotations \cite{skerry2018towards, wang2018style}. Specifically, they use neural networks to learn prosody embeddings from the reference speech in an unsupervised manner. These methods have demonstrated their ability to generate speech with expressive styles. Inspired by their success, we propose to use prosody embeddings to model prosody in the PPGs based VC frameworks.

Secondly, vocoders influence the quality of the converted speech. Prior works utilized parametric vocoders (such as STRAIGHT \cite{kawahara1999restructuring} and WORLD \cite{morise2016world}). However, these vocoders limit the quality of converted speech. To deal with this problem, researches focus on neural vocoders \cite{oord2016wavenet, kalchbrenner2018efficient}. WaveNet \cite{oord2016wavenet} is one of the state-of-the-art neural vocoders. It predicts every sample by the previous observations without much assumptions based on the prior knowledge related with speech. However, since WaveNet relies on sequential generation of one audio sample at a time, it is hard to deploy in a real-time production setting. Recently, an efficient neural vocoder called LPCNet \cite{valin2019lpcnet} is proposed. LPCNet combines linear prediction with recurrent neural network and directly predicts the excitation rather than the sample values. Compared with WaveNet, LPCNet can generate speech in real time. Meanwhile, since LPCNet depends directly on the linear predictive coding filter shape, it can better control over the outputs of the spectral shape. Therefore, we apply LPCNet vocoder for speech generation.

In this paper, we propose a new PPGs based VC framework for any-to-one voice conversion. The main contributions of this paper lie in three aspects: 1) To better control the prosody of converted speech, we propose to learn prosody embeddings from the source speech in an unsupervised manner; 2) To synthesize speech with close to natural quality in real time, we propose to use the LPCNet vocoder for speech generation; 3) Experimental results on the popular benchmark datasets CMU-ARCTIC and our Mandarin corpus demonstrate the effectiveness of our proposed method. We find that our method can improve the speech quality and speaker similarity of the converted speech. Meanwhile, our method can even achieve promising results when the source speech is a singing song.

The remainder of this paper is organized as follows: In Section 2, we describe our baseline voice conversion system. Section 3 presents our proposed VC framework with prosody embeddings. Section 4 illustrates the experimental data, setup, results and discussion in detail. Finally, we give a conclusion of the proposed work in Section 5.

\section{Baseline system}
We present our baseline voice conversion approach using the pitch for prosody control and LPCNet for speech synthesis.

\begin{figure*}[t]
	\centering
	\includegraphics[width=\linewidth]{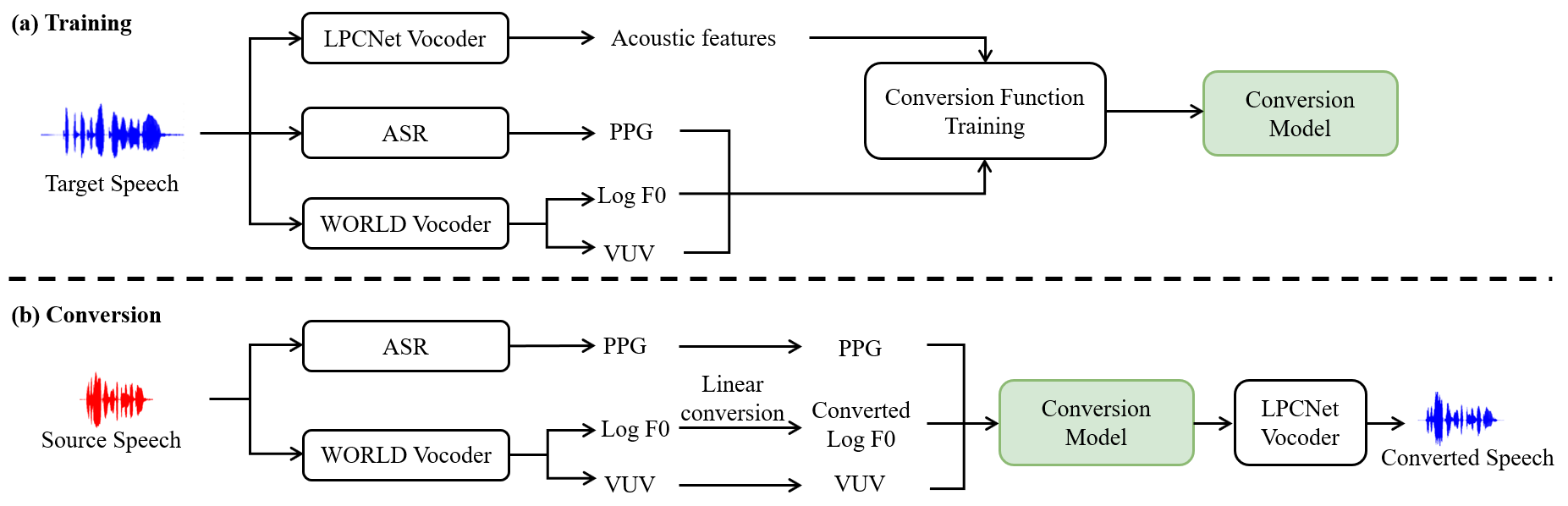}
	\caption{The framework of the baseline system.}
\end{figure*}

\subsection{Linear Prediction Coding Net (LPCNet)}
LPCNet \cite{valin2019lpcnet} is a WaveRNN \cite{kalchbrenner2018efficient} variant that uses the neural networks to generate speech samples from Bark-Frequency Cepstral Coefficients (BFCCs) \cite{moore2012introduction}, pitch period and pitch correlation parameters. In this work, we use the code published by the Mozilla team \cite{valin2019lpcnet} with some modifications. To better control high frequency features, we increase 18-dimensional BFCCs to 30-dimensional BFCCs. What's more, we utilize the OpenBLAS toolkit to accelerate the LPCNet inference. Therefore, our acoustic features contains 30-dimensional BFCCs, 1-dimensional pitch period and 1-dimensional pitch correlation.

\subsection{Framework Overview}
In the training stage (Figure 1(a)), we extract the acoustic features $Y \in \mathbb{R} ^{T \times D_{a}}$, PPGs $L \in \mathbb{R} ^{T \times D_p}$, pitch $f0 \in \mathbb{R} ^{T \times 1}$ and voiced/unvoiced flag (vuv) $f_{vuv} \in \mathbb{R} ^{T \times 1}$ from the given speech data of the target speaker, where $T$ is the number of frames. $D_a$ and $D_p$ represent the feature dimension of the acoustic features and PPGs, respectively. To control the prosody of generated speech, we form the input features by concatenating the pitch, vuv and PPGs, denoted as $F=[L;f0;f_{vuv}] \in \mathbb{R} ^{T \times (D_p+2)}$. Then a CBHG \cite{wang2017tacotron} conversion model is trained to map the input $F$ to the output acoustic features. Concretely, the CBHG \cite{wang2017tacotron} conversion model consists of a bank of 1-D convolutional filters, followed by highway networks and a bidirectional GRU.

In the conversion stage (Figure 1(b)), we first extract the PPGs, pitch and vuv features from the source speech. As the feature mismatch exists between the source speaker's pitch and the target speaker's pitch, a linear conversion is applied:

\begin{equation}
log f0_y = \left( log f0_x-\mu_x \right) * \left( \frac{\sigma_y}{\sigma_x} \right) + \mu_y
\end{equation}
where $\mu_x$ and $\sigma_x$ are the mean and variance of the source speaker’s log $f0$, respectively. $\mu_y$ and $\sigma_y$ are the mean and variance of the target speaker’s log $f0$. $log f0_x$ and $log f0_y$ are the source and converted $f0$ in logarithmic domain, respectively. Then we concatenate the PPGs, converted pitch and vuv features together. These representations are used as the input to the conversion model to predict the converted acoustic features. Finally, we feed these converted acoustic features to the LPCNet vocoder for speech generation.

\subsection{Limitations}
Despite the good performance of the baseline system, it still has some limitations. Firstly, the pitch detection algorithms face challenges in some challenging situations (e.g., singing conditions \cite{yeh2012hybrid}). Secondly, the prosody is related with many factors, including the intonation, stress, rhythm and pitch. However, we only utilize the pitch to control the prosody of the generated speech, thus limiting the speech quality of converted speech.

\begin{figure*}[h]
	\centering
	\includegraphics[width=\linewidth]{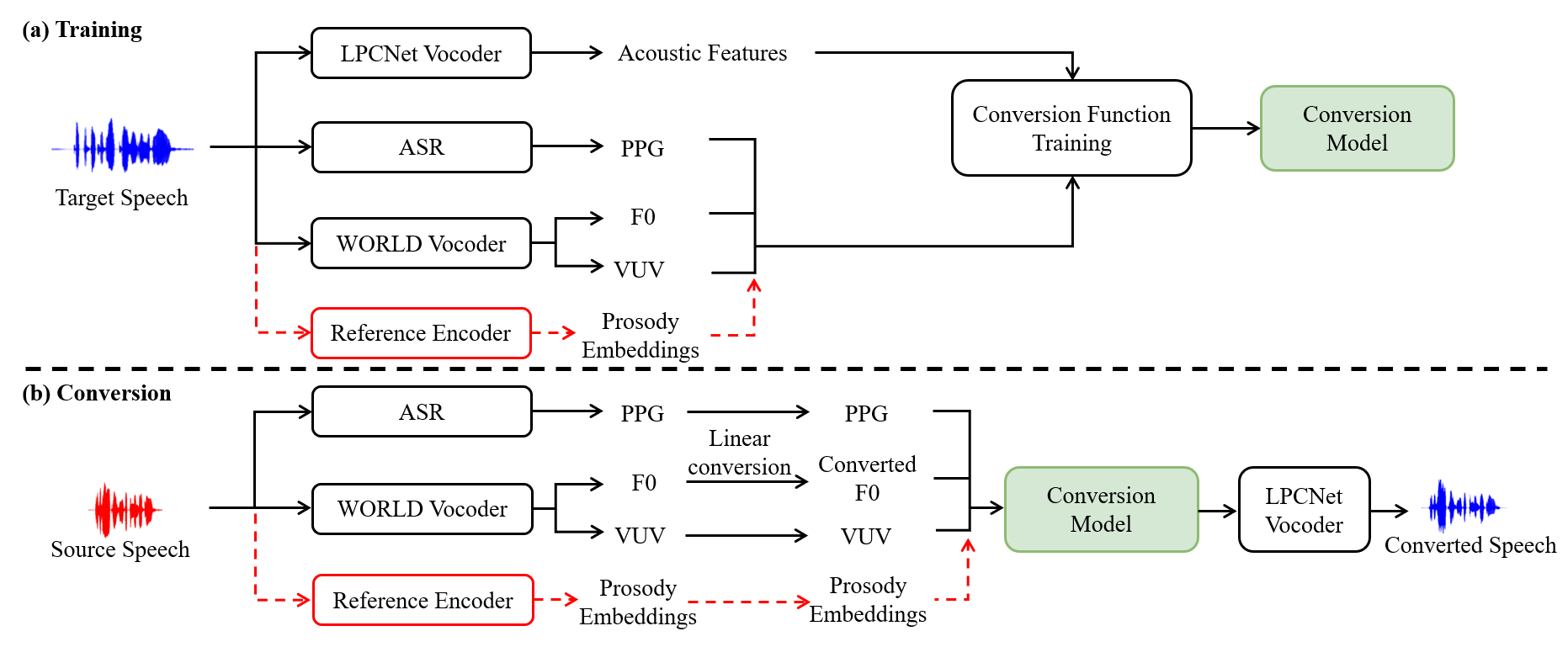}
	\caption{The framework of the proposed system.}
\end{figure*}

\section{Proposed Method}
To overcome above limitations, we utilize the reference encoder \cite{skerry2018towards} to learn prosody embeddings from the input speech. The reference encoder is plugged into the baseline system and trained without any other supervision except for the VC's reconstruction error.

\subsection{Framework Overview}
During training (Figure 2(a)), different from the baseline system, we also learn prosody embeddings $P \in \mathbb{R} ^{T \times D_e}$ from the reference encoder, where $D_e$ represents the feature dimension of prosody embeddings. To control the prosody of converted speech, we form the input features by concatenating the pitch $f0 \in \mathbb{R} ^{T \times 1}$, voiced/unvoiced flag $f_{vuv} \in \mathbb{R} ^{T \times 1}$, PPGs $L \in \mathbb{R} ^{T \times D_p}$ and prosody embeddings $P \in \mathbb{R} ^{T \times D_e}$, denoted as $F=[L;f0;f_{vuv};P]$, where $F \in \mathbb{R} ^{T \times (D_p+D_e+2)}$. Then, a CBHG \cite{wang2017tacotron} conversion model is trained to map the input $F$ to the output acoustic features.

At run-time (Figure 2(b)), we first extract the pitch, vuv, PPGs and prosody embeddings for an arbitrary source speech. Then we transform the pitch of the source speaker into that of the target speaker by the linear conversion in Eq. (1). We concatenate these representations as the input, and transform the input to acoustic features by the conversion model. Finally, we use LPCNet for speech generation.

\begin{figure}[t]
	\centering
	\includegraphics[width=0.55\linewidth]{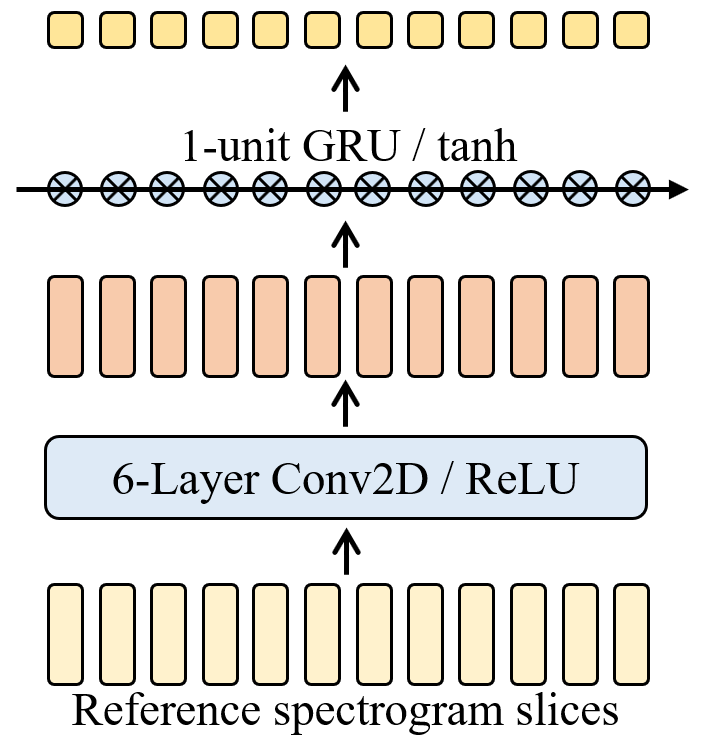}
	\caption{The prosody reference encoder module. A 6-layer stack of 2D convolutions with ReLU activations, followed by a single-layer GRU with 1 unit and a \emph{tanh} activation.}
\end{figure}

\subsection{Reference Encoder}
Speech is encoded to prosody embeddings using the reference encoder. Earlier works focused on fixed-length prosody embeddings regardless of the length of the input speech \cite{skerry2018towards}. These prosody embeddings lose the temporal information. However, the temporal information is an important aspect of prosody. Therefore, we focus on variable-length prosody embeddings.

As shown in Figure 3, the reference encoder takes a mel-spectrogram $I \in \mathbb{R} ^{T \times D_m}$ as the input, where $T$ is the length of the mel-spectrogram and $D_m$ is the feature dimension. This network contains 6-layer 2D-convolutional layers. Each layer is composed of 3$\times$3 filters with 1$\times$2 stride, SAME padding and ReLU activation. The number of filters in each layer are [32, 32, 64, 64, 128, 128]. The outputs of the last convolutional layer is fed to a uni-directional Gated Recurrent Unit (GRU) with one unit and \emph{tanh} activation. The outputs of GRU at every timestep form the variable-length prosody embeddings $P \in \mathbb{R} ^{T \times D_e}$.

Meanwhile, we also test following modifications in the reference encoder: 1) CoordConv \cite{liu2018intriguing} can augment the positional information to the input. As positional information is useful to encode prosody sequentially \cite{lee2019robust}, we use CoordConv for the first convolutional layer. 2) Bi-directional GRU can get the information from frames occurring before and after itself in the mel-spectrogram. As the contextual information is useful to encode prosody, we replace uni-directional GRU with bi-directional GRU. However, we do not observe performance improvement with these modifications. Therefore, we utilize the reference encoder in Figure 3 to extract prosody embeddings.

\section{Experiments and Discussion}
In this section, we first present our experimental databases. Then, we illustrate implementation details of our proposed method and several baseline models. Finally, we compare our method with baseline approaches via subjective measures. Speech samples from the following experiments are available online at \href{https://zeroqiaoba.github.io/voice-conversion}{https://zeroqiaoba.github.io/voice-conversion}.

\subsection{Corpus Description}
Our experiments are conducted on two databases: 1) CMU-ARCTIC American English database \cite{kominek2003cmu}; 2) Our Mandarin Voice Conversion database. The CMU ARCTIC database consists of 4 native American English speakers (2 male BDL and RMS, 2 female CLB and SLT), and each speaker has 1,134 sentences. Intra-gender and inter-gender conversions are conducted between following pairs: CLB to BDL (F2M), CLB to SLT (F2F), RMS to BDL (M2M) and RMS to SLT (M2F). Our Mandarin Voice Conversion database contains 2 native Mandarin Chinese speakers (1 female TS, 1 male SONG). TS has 15000 sentences and SONG has 500 sentences. Our experiments are conducted between SONG to TS (M2F). 

To increase the amount of training samples, we perform data augmentation by means of speed perturbation. Speech perturbation is a technique of changing speech speed without the tone changed. It is performed on original signals with speed factor 0.4, 0.6, 0.8, 1.0 and 1.2. In our voice conversion experiments, we use 1,000 sentences for training, and another 20 non-overlap utterances of each speaker are used for evaluation.

\begin{figure*}[t]
	\centering
	\subfigure[MOS test results with 95\% confidence intervals to assess speech quality]{
		\begin{minipage}[t]{0.5\linewidth}
			\centering
			\includegraphics[width=3.3in]{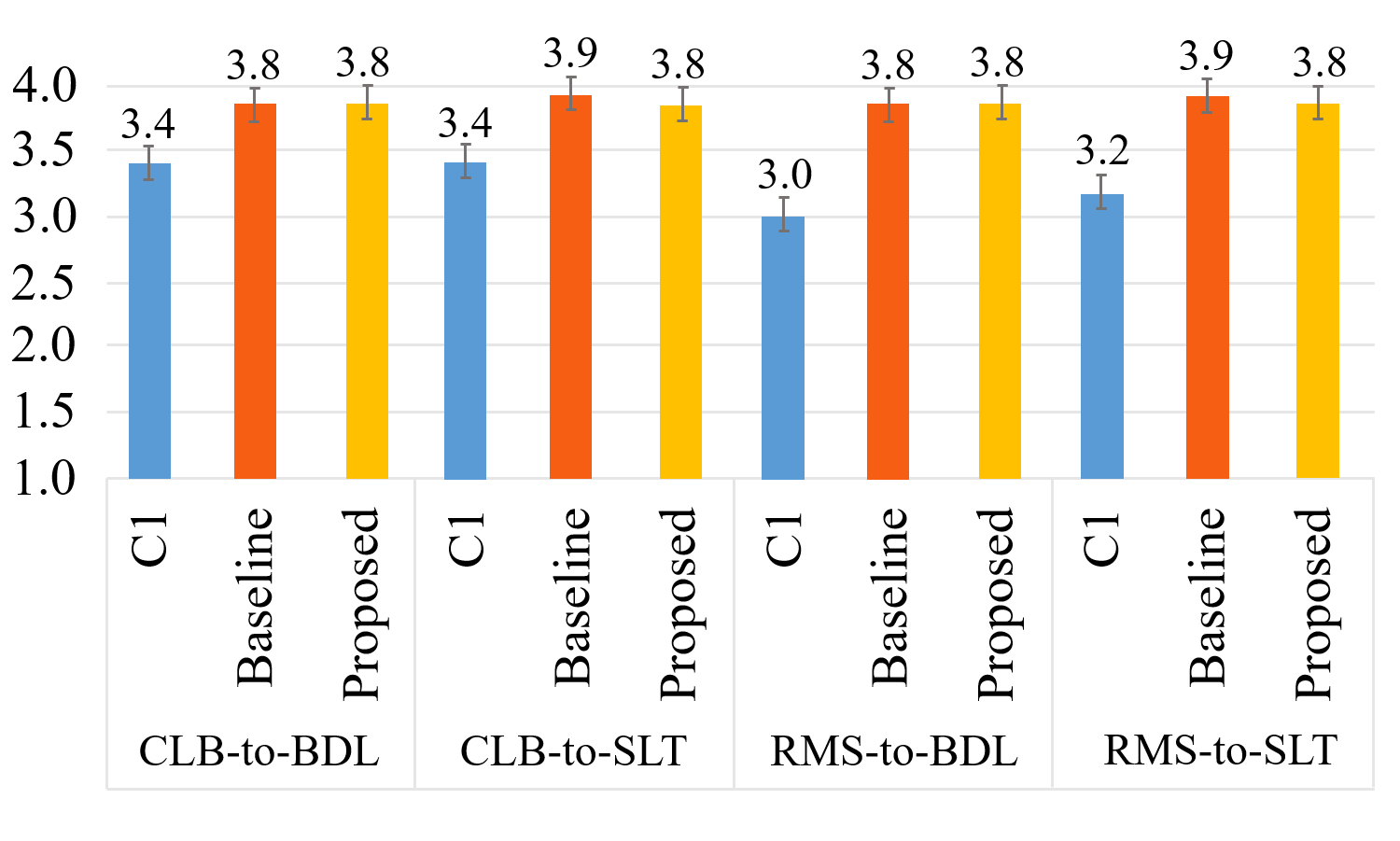}
			\label{fig4a}
		\end{minipage}%
	}%
	\subfigure[Same/Different paradigm to assess speaker similarity]{
		\begin{minipage}[t]{0.5\linewidth}
			\centering
			\includegraphics[width=3.3in]{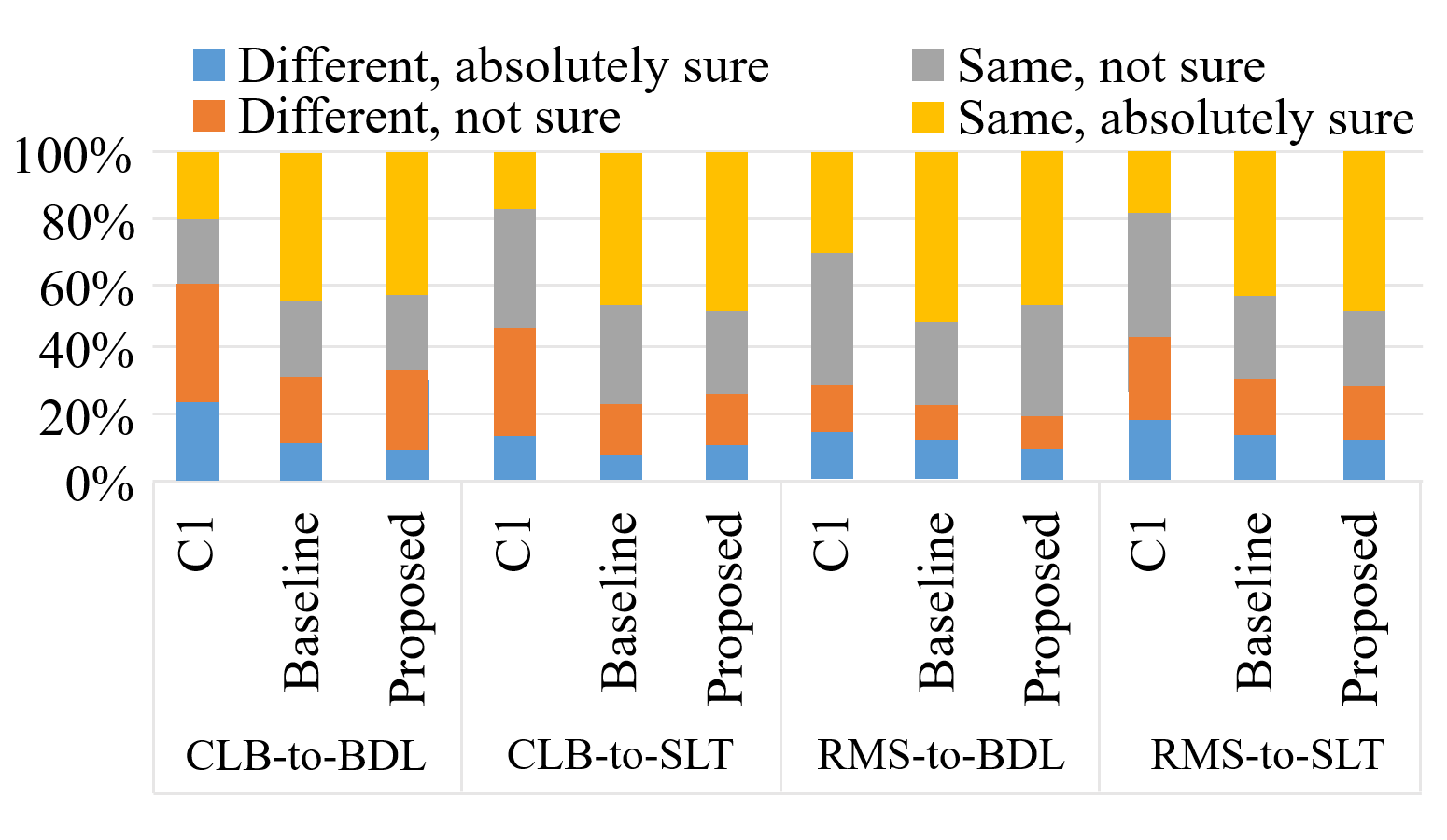}
			\label{fig4b}
		\end{minipage}%
	}%
	\centering
	\caption{Subjective test results on the CMU-ARCTIC American English database.}
	\label{fig4}
\end{figure*}

\begin{figure}[t]
	\centering
	\subfigure[MOS test results with 95\% confidence intervals]{
		\label{fig5a}
		\includegraphics[width=\columnwidth]{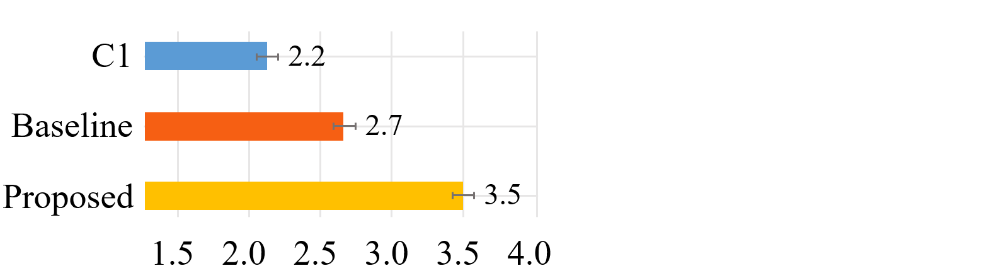}
	}
	\subfigure[Same/Different paradigm to assess speaker similarity]{
		\label{fig5b}
		\includegraphics[width=\columnwidth]{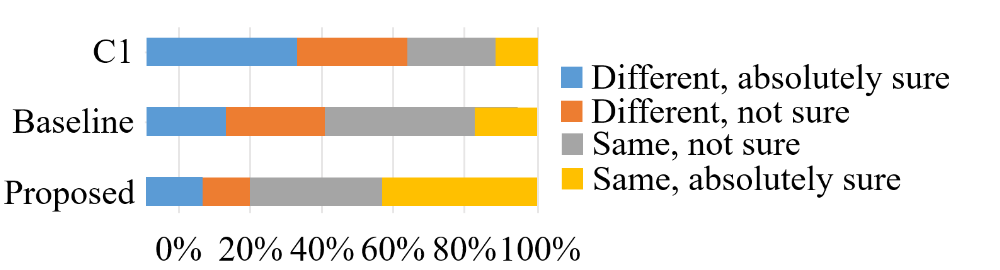}
	}
	\caption{Subjective test results on our Mandarin Voice Conversion database (for the SONG-to-TS pair).}
	\label{fig5}
\end{figure}

\subsection{Implementation Details}
All features are extracted with 10ms window shift. WORLD \cite{morise2016world} is used to extract 1-dimensional pitch and 1-dimensional vuv. LPCNet \cite{valin2019lpcnet} is used to extract 32-dimensional acoustic features, including 30-dimensional BFCCs, 1-dimensional pitch period and 1-dimensional pitch correlation. The librosa toolkit \cite{mcfee2015librosa} is used to extract 80-dimensional mel-spectrograms. The 512-dimensional PPGs are extracted from the acoustic model in SI-ASR, which is implemented using the Kaldi toolkit \cite{povey2011kaldi} and trained on our 20,000 hours corpus.

The CBHG \cite{wang2017tacotron} conversion model has a bank of 1-D convolutional filters, followed by highway networks (4 layers with 64 hidden units) and a bidirectional GRU (64 units for each GRU component). Concretely, it has $K=16$ sets of 1-D convolutional filters, where the $k$-th set contains 128 filters of width $k$ ($k \in [1, K]$). As variable-length prosody embeddings have a large enough capacity to copy the input speech, we use a very small dimension of bottleneck size. In this paper, we use 1-dimensional prosody embeddings. To optimize the parameters, we use the Adam optimizer with a learning rate of 0.001. We train our models for at least 100k steps with a batch size of 32. Gradient clipping is also used for regularization.

Three systems are evaluated in the experiments. In additional to the baseline system ($\textbf{\emph{{Baseline}}}$) and the proposed system ($\textbf{\emph{{Proposed}}}$), another comparison system is also implemented to verify the effectiveness of our proposed method. Comparison system 1 ($\textbf{\emph{{C1}}}$) \cite{tian2018average}: It employs PPGs based mapping approach to voice conversion without using the parallel training data. Firstly, it trains a multi-speaker average model that maps PPGs to speaker-dependent acoustic features. Then, it adapts a pre-trained multi-speaker model for the target speaker. Finally, the WORLD vocoder \cite{morise2016world} is utilized to generate the converted waveform.

\subsection{Subjective Evaluation}
Followed with previous works \cite{zhou2019cross, tian2019speaker}, the quality of the speech samples and their similarity to the target speaker are evaluated using the subjective evaluation. The Mean Opinion Score (MOS) tests are conducted to assess speech quality. In the MOS tests, listeners are asked to rate the converted speech on a 5-point scale, ranging from 1 (completely unnatural) and 5 (completely natural). Meanwhile, we conduct the Same/Different paradigm to assess speaker similarity. In this test, the listeners are asked to compare and select whether the converted samples are uttered by the same target speaker. In practice, 12 subjects with normal hearing participate in all tests. 20 utterances in the test set for each conversion pair are randomly selected and converted using our proposed method and two baseline methods. The listeners are asked to use headphones and samples are shown to them in the random order.

Subjective test results on the CMU-ARCTIC American English database and our Mandarin Voice Conversion dataset are shown in Figure \ref{fig4} and Figure \ref{fig5}, respectively. As can be seen, $\textbf{\emph{{C1}}}$ achieves worse performance in all conversion pairs. These results suggest that LPCNet outperforms WORLD in terms of both speech quality and speaker similarity. Meanwhile, we observe that the proposed method significantly outperforms $\textbf{\emph{{Baseline}}}$ for the $\emph{SONG-to-TS}$ pair. As for other conversion pairs in the CMU-ARCTIC database, we observe that the performance of $\textbf{\emph{{Baseline}}}$ is comparable to that of $\textbf{\emph{{Proposed}}}$. We find the reason lies in two aspects. Firstly, the pitch detection algorithm makes some errors in the singing conditions. Secondly, the pitch cannot describe the prosody of the singing songs perfectly. Since our method enables fine-grained control the prosody of the generated speech via prosody embeddings, we achieve better performance than $\textbf{\emph{{Baseline}}}$.

\section{Conclusions}
In this paper, we propose a voice conversion framework based on prosody embeddings and LPCNet. The prosody embeddings enable fine-grained control of the prosody of generated speech. LPCNet can synthesize speech with close to natural quality while running in real time. Subjective evaluations show that the proposed method can achieve both high naturalness and high speaker similarity in singing conditions.

\bibliographystyle{IEEEtran}
\bibliography{mybib}

\end{document}